\documentclass[twoside,a4paper,10pt]{article}

\usepackage[english,brazil]{babel}

\usepackage[utf8]{inputenc}
\usepackage[T1]{fontenc} 

\newcommand{\shorttitle}{Non-binary artificial neuron implemented on a quantum computer}

\usepackage{authblk} 
\usepackage{lipsum} 
\usepackage{times}
\usepackage{microtype} 
\usepackage[hmarginratio=1:1,top=33mm,columnsep=20pt,text={18cm,24.3cm}]{geometry} 
\usepackage{booktabs} 
\usepackage{abstract} 
\newcommand{\titleeng}{\vspace{4mm}\fontsize{14pt}{14pt}
\selectfont}
\newcommand{\titlept}{\textbf} 
\usepackage{titlesec} 
\usepackage{fancyhdr} 
\usepackage{amssymb,amsmath,amsthm,amsfonts} 
\usepackage{natbib} 
\usepackage{graphicx,rotating} 
\usepackage{enumerate}
\usepackage{graphics}
\usepackage[dvips]{epsfig}
\usepackage{multirow}
\usepackage{nonfloat}
\usepackage{multicol} 
\usepackage{url} 
\usepackage{hyperref} 
\usepackage{subfigure} 
\usepackage{bm}
\usepackage{color}
\usepackage{icomma} 

\allowdisplaybreaks[4] 

\title{
\titlept{Neurônio artificial não-binário com variação de fase implementado em um computador quântico}
\\ 
\titleeng{Non-binary artificial neuron with phase variation implemented on a  quantum computer} 
} 

\author{Jhordan Silveira de Borba¹ and Jonas Maziero²} 
 
\date{1 - Universidade Federal do Rio
Grande do Sul (UFRGS), Porto
Alegre, Rio Grande do Sul, Brasil. \\ 2 - Universidade Federal de Santa
Maria (UFSM), Santa Maria, Rio
Grande do Sul, Brasi}
 
\begin{document}

\maketitle 

\thispagestyle{empty} 

\begin{abstract}

\noindent O primeiros neur\^onios artificiais qu\^anticos seguiram um caminho similar aos modelos cl\'assicos, envolvendo apenas estados discretos. Neste artigo, introduzimos um algoritmo que generaliza o modelo bin\'ario, manipulando a fase de n\'umeros complexos. Propomos, testamos e implementamos um modelo de neur\^onio que envolve valores cont\'inuos em um computador qu\^antico. Atrav\'es de simula\c{c}\~oes, demonstramos que nosso modelo funciona em um esquema h\'ibrido utilizando o gradiente descendente como algoritmo de aprendizado. Esse trabalho representa um novo passo na dire\c{c}\~ao da avalia\c{c}\~ao do uso de redes neurais implementadas eficientemente em dispositivos qu\^anticos de curto-prazo.
\smallskip \\
\noindent \textbf{Palavras-chave:} Computa\c{c}\~ao qu\^antica,  Mec\^anica qu\^antica, Rede neural artificial, Neur\^onio sigmoide, Gradiente descendente.

\end{abstract}

{
\selectlanguage{english}
\begin{abstract}

\noindent The first artificial quantum neuron models followed a similar path to classic models, as they work only with discrete values. Here we introduce an algorithm that generalizes the binary model manipulating the phase of complex numbers. We propose, test, and implement a neuron model that works with continuous values in a quantum computer. Through simulations, we demonstrate that our model may work in a hybrid training scheme utilizing gradient descent as a learning algorithm. This work represents another step in the direction of evaluation of the use of artificial neural networks efficiently implemented on near-term quantum devices.


\end{abstract}
}

\newpage


\section{Introduction}

Artificial neural networks constitute a class of computational models that achieved a high success rate at specific tasks, as for example in pattern recognition. Although, in modern practical applications, artificial neural networks run usually as classic algorithms in conventional computers \cite{Tacchino}, there is a rising interest in their application in quantum devices. Intrinsic properties of quantum mechanics such as the storage and representation of large data quantities as vectors and matrices provide exponential growth in both the storage and processing power \cite{Feldman,Plank,Dharmendra,Solano}.

An artificial neuron model  implemented in a current quantum computer was published in 2019 \cite{Tacchino}. This model is based on the Rosenblatt neurons, limited to binary values $i_{j},w_{j}\in\left\{ -1,1\right\} $ and utilized a learning rule based in hyper-planes. A possible improvement from this model is to extend it to continuous values.

In the literature about models of artificial neurons with continuous values, the sigmoid neuron is recognized as one of the main models implemented in modern neural networks. On the other hand, the gradient descent is the standard learning algorithm \cite{Livro_Online}.  Furthermore, in the literature about quantum neuron models, inspired by network models of biologic neurons in which the neurons are sensible not only to the amplitude of input signals but also to its phase, it was proposed to consider the phase of continuous values to encode the information, beyond the amplitude of the complex numbers \cite{Altaisky, Mitja}. Sutherland even interprets the phase like ``real'' information and the amplitude as the level of confidence of the correspondent information \cite{Sutherland}.

So, our work proposes modifications in the binary model exploring properties from complex numbers that are inherent to quantum mechanics. Firstly, we encode the classic \textit{m}-dimensions input vector in the quantum hardware using \textit{N} qubits, on which $m=2^N$, exploring the exponential advantage of quantum information storage. Secondly, we construct the quantum circuit using two different algorithms: first implementing a procedure directly by `brute-force' through a phase rotation block, and then implementing an adaptation of the \textit{Hypergraph States Generation Subroutine}  (HSGS)  algorithm.

Experimentally, we implement a two-qubit version of the algorithms in the IBM processors available for cloud computing. We demonstrate that the proposed algorithm provides the results expected in classic approaches, as well as the optimization expected from HSGS algorithm if compared to the phase rotation block. We report the potential of our neuron model, we also simulated a training scheme to recognize simple patterns with 2+1 qubits.

Therefore, this work provides an important contribution to the efficient implementation of machine learning applications through both quantum processing and quantum devices.


\section{Theory}
\subsection{Modeling the quantum circuit from neuron}
An input vector with $m$ dimensions is encoded using $m$ coefficients to define a general function wave $\left|\psi_{i}\right\rangle $. In the practice, for  any arbitrary inputs and weights:

$$\overrightarrow{i}=\left(\begin{array}{c}
i_{0}\\
i_{1}\\
\vdots\\
i_{m-1}
\end{array}\right),\ \qquad\overrightarrow{w}=\left(\begin{array}{c}
w_{0}\\
w_{1}\\
\vdots\\
w_{m-1}
\end{array}\right)$$
Then the quantum states may be defined as follows:
$$\left|\psi_{i}\right\rangle =\frac{1}{\sqrt{m}}\sum_{j=0}^{m-1}i_{j}\left|j\right\rangle,\  \qquad\left|\psi_{w}\right\rangle =\frac{1}{\sqrt{m}}\sum_{j=0}^{m-1}w_{j}\left|j\right\rangle  $$

\begin{figure}[h]
\centering
\includegraphics[width=0.4\textwidth]{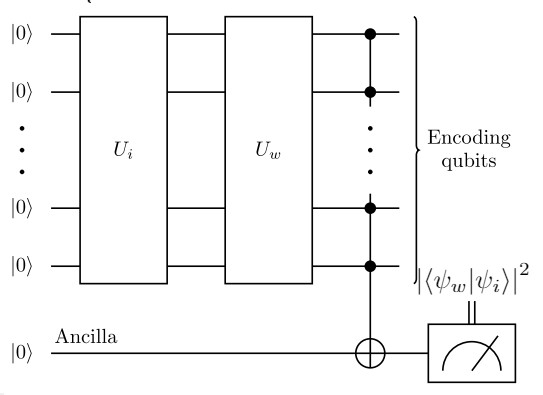}
\caption{General scheme of the quantum algorithm that calculates the inner product.}\label{UM}
\end{figure}

The states $\left|j\right\rangle \in\left\{ \left|00\ldots00\right\rangle ,\left|00\ldots01\right\rangle ,\dots,\left|11\ldots11\right\rangle \right\} $ constitute a basis, called computational basis, from the quantum processor and may be identified by decimal integers that are made by the respective binary string $\left|j\right\rangle \in\left\{ \left|0\right\rangle ,\left|1\right\rangle ,\dots,\left|m-1\right\rangle \right\} $. Evidently, if we have $n$ qubits in the register, so we have $m=2^{N}$ basis states. Moreover, the information may be encoded in a uniform superposition of the entire computational basis since we may change the phase of complex numbers without changing its modulus. Hence, the present work has as its initial objective to develop a quantum circuit able to perform the following calculation:
$$\left\langle \psi_{i},\psi_{w}\right\rangle =\frac{\overrightarrow{w}\cdot\overrightarrow{v}}{m} $$
If we assume that qubits are initialized in the state $\left|00\dots00\right\rangle \equiv\left|0\right\rangle ^{\otimes N}$, the first step is to prepare the state $\left|\psi_{i}\right\rangle $ through a unitary transformation $U_{i}$ with the input values encoded at $\overrightarrow{i}$. That is, the input operator must be able to do the following transformation:
$$U_{i}\left|0\right\rangle ^{\otimes N}=\left|\psi_{i}\right\rangle $$
In principle, any unitary matrix $m \times m$ with $\overrightarrow{i}$ in the first column works. The next step is to do a unitary transformation $U_{w}$ with the weight values $\overrightarrow{w}$ encoded in some way that the result of the inner product finishes in the last coefficient of the computational basis. Explicitly:
\begin{equation}
\begin{split}
U_{w}\left|\psi_{i}\right\rangle & =\left[\frac{1}{\sqrt{m}}\left(\begin{array}{cccc}
A_{11} & A_{21} & \cdots & A_{1\left(m-1\right)}\\
A_{12} & A_{22} & \cdots & A_{2\left(m-1\right)}\\
\vdots & \vdots & \ddots & \vdots\\
\overline{w_{0}} & \overline{w_{1}} & \ldots & \overline{w_{m-1}}
\end{array}\right)\right]\left[\left(\begin{array}{c}
i_{0}\\
i_{1}\\
\vdots\\
i_{m-1}
\end{array}\right)\frac{1}{\sqrt{m}}\right] \\
& = \left(\begin{array}{c}
C_{1}\\
C_{2}\\
\vdots\\
\sum_{j=0}^{m-1}\overline{w_{j}}i_{j}
\end{array}\right)\frac{1}{m}
\end{split}
\end{equation}
Where $A_{ij}$ and $C_{i}$ denote irrelevant coefficients to our purpose. Similarly to the case of the input operator, in principle any unitary matrix $m \times m$ with the conjugate of $\overrightarrow{w}$ in the last row works. In order to demonstrate the feasibility of the proposed algorithms, the following sections encompass some demonstrations. Especially, the operator $U_{w}$ must be able to perform the following transformation:
$$U_{w}\left|\psi_{w}\right\rangle =\left|1\right\rangle ^{\otimes N}=\left|m-1\right\rangle$$
In this way, after the application of both operators, the final state may be decomposed in the computational basis in the following way:
$$U_{w}U_{i}\left|0\right\rangle ^{\otimes N}=U_{w}\left|\psi_{i}\right\rangle =\sum_{j=0}^{m-1}C_{j}\left|j\right\rangle $$
Denoting the final state as $\left|\phi_{i,w}\right\rangle \equiv U_{w}U_{i}\left|0\right\rangle  ^{\otimes N}$, and like the $U_{w}$ is a unitary operator:
\begin{equation}
\begin{split}
\left\langle \psi_{w} | \psi_{i} \right\rangle &  =\left\langle \psi_{w} \left| U_{w}^{\dagger}U_{w} \right| \psi_{i} \right\rangle \\
& =\left\langle m-1|\phi_{i,w}\right\rangle \\
& =\sum_{j}^{m-1}C_{j}\left\langle m-1|j\right\rangle \\
& =C_{m-1}
\end{split}
\end{equation}It is evident that we aim to acquire the coefficient $C_{m-1} $ of the final state after the application of both unitary transformations. We may also use an auxiliary qubit $a$ initially in the state $\left|0\right\rangle $ and a multi controlled NOT gate among the $N$ qubits that encode the information and the target $a$, which leads to:
$$\left|\phi_{i,w}\right\rangle \left|0\right\rangle \rightarrow\sum_{j=0}^{m-2}c_{j}\left|j\right\rangle \left|0\right\rangle _{a}+c_{m-1}\left|-1\right\rangle \left|1\right\rangle _{a}$$
Thus, it is only necessary to measure the probability of auxiliary qubit $a$ being at state $\left|1\right\rangle$. Concluding, the final output from our quantum circuit will be the value measured $\left|C_{m-1}\right|^{2}$, or yet:
$$\left|\left\langle \psi_{w}|\psi_{i}\right\rangle \right|^{2}=\frac{\left|\overrightarrow{w}\cdot\overrightarrow{v}\right|^{2}}{m^{2}}$$ 
Figure \ref{UM} illustrates a general scheme of the algorithm that performs the aforementioned calculus.

\subsection{Implementation of unitary transformations}
\subsubsection{Phase rotation block}

Firstly, we define the phase rotation block $B_{N,j}\left(\lambda\right)$ as a unitary transformation acting in the computational basis of $N$ qubits as follows:

$$B_{N,j} \left( \lambda \right) \left| j' \right\rangle = \left\{ \begin{array}{ccc} e^{i\lambda} \left| j' \right\rangle & ,\text{se} & j=j' \\ \left| j' \right\rangle & , \text{se} & j\neq j' \end{array} \right. $$
We may implement a family of rotation blocks for $N$ qubits through a multi controlled phase rotation $C_{u1}\left(\lambda\right)$ with NOT operators in single qubits
$$B_{N,j}\left(\lambda\right)=O_{j}\left[C_{u1}\left(\lambda\right)\right]O_{j}$$
Where
$$O_{j}=\otimes_{l=0}^{N}\left(NOT_{l}\right)^{1-j_{l}}$$
This expression is responsible for taking the target coefficient to state $\left|m-1\right\rangle $ from the computational basis. Then it is possible to apply the multi controlled rotation of phase $C_{u1}\left(\lambda\right)$  with the $N$ qubit as the target. In this way we will only change the target coefficient. After this procedure, $O_j$ is once more utilized in order to return the  coefficient to its original state.

$NOT_{l}$ means that we applied the operator $NOT$ in the $l$-th qubit, and $j_{l}=0$ if the $l$-th qubit is in the state $\left|0\right\rangle $ ( $j_{l}=1$ if the state is $\left|1\right\rangle $). It is important to notice that $j_{l}=0$ imply in the fact that we will apply the operator $NOT$, but $j_{l}=1$ means that we will use an identity operator to maintain the $l$-th qubit unaltered. It is also important to notice that $\otimes$ denote a tensor product.

With the rotation block defined, we may summarize the complete sequence for implementation of $U_{i}$, remembering from the definition of input operator ($U_{i}\left|0\right\rangle ^{\otimes N}=\left|\psi_{i}\right\rangle $), in the following way:
\begin{itemize}
\item Starting from the register initialized as $\left|0\right\rangle ^{\otimes N}$, we apply the parallel Hadamard (H) operators in all qubits to create a state of total superposition between all elements from the computational basis:$\left|0\right\rangle ^{\otimes N}\stackrel{H^{\otimes N}}{\rightarrow}\frac{1}{\sqrt{m}}\sum_{j=0}^{m-1}\left|j\right\rangle \equiv\left|\psi_{0}\right\rangle  $
\end{itemize}
\begin{itemize}
\item Then we need to apply the rotation blocks in the coefficients that are needed.
\begin{itemize}
\item Since the rotation block keeps the other coefficients unaltered, the sequence of application is irrelevant.
\end{itemize}
\end{itemize}
The operator $U_{w}$ is defined similarly. However, we need to remember its definition ( $U_{w}\left|\psi_{w}\right\rangle =\left|1\right\rangle ^{\otimes N}=\left|m-1\right\rangle$):
\begin{itemize}
\item The rotation blocks are applied with the negative phase from the weights. This process may  provide a superposition of the computational basis state: $\left|\psi_{w}\right\rangle \stackrel{B_{N,j}\left(- \lambda\right)}{\rightarrow}\left|\psi_{0}\right\rangle $
\item After employing the Hadamard operator in all qubits, the system final state corresponds to the processor initial state ($\left|\psi_{0}\right\rangle \stackrel{H^{\otimes N}}{\rightarrow}\left|0\right\rangle ^{\otimes N}$)
\item Finally, after applying the NOT operator in all qubits, the system state reaches the desired state: $\left|0\right\rangle ^{\otimes N}\stackrel{NOT^{\otimes N}}{\rightarrow}\left|1\right\rangle ^{\otimes N}$
\end{itemize}

\begin{figure}[h]
\includegraphics[width=1\textwidth]{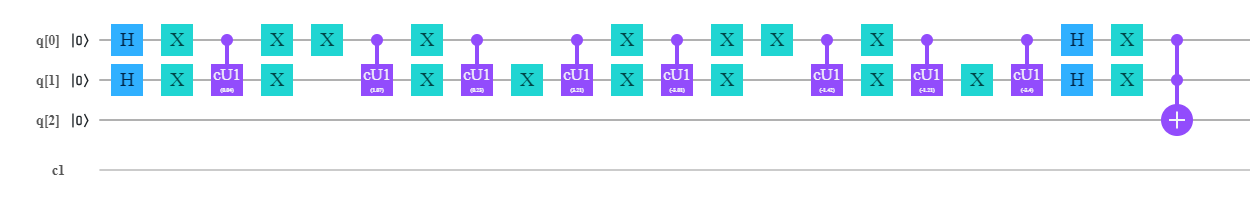}
\centering
\caption{Example of a circuit utilizing the rotation blocks algorithm.}
\label{DOIS}
\end{figure}

It is interesting to notice that the complex conjugate of any complex number ($z=e^{i\lambda}$) is equivalent to another complex number with the same module but with a phase that is equivalent to its negative ($\overline{z}=e^{-i\lambda}$).  Therefore, multiplying $z$ by its conjugate is equivalent to applying a phase rotation with the negative of your phase: $\overline{z}z=e^{-i\lambda}e^{i\lambda}=1$. Thus, we are building an operator $U_{w}$ whose last row is the conjugate from $\overrightarrow{w}$, as we desire. Figure \ref{DOIS} illustrates an example of a circuit built utilizing rotation blocks.

\subsubsection{HSGS}

A more efficient solution is given by the algorithm called HSGS. This algorithm reduces the needed quantum resources in comparison to rotation blocks. This occurs mainly because even if the HSGS algorithm keeps showing an exponential cost, it optimizes the required multi controlled operators \cite{Tacchino}.

Working first with the construction of the input operator $U_{i}$, starting from the total superposition state obtained via $H^{\otimes N}$, the first step is to apply the operator $u_{1}$ to the coefficients of the states that have only $p=1$ bit in the state $\left|1\right\rangle $ (for example $\left|0\dots010\dots0\right\rangle $). Then we apply the required corrections to the states that show $p=2$ using the multi controlled phase rotation $c^{p}u_{1}$ until $p=N$. Only $\left|0\right\rangle ^{\otimes N}$  remains unaltered. If needed, we may utilize $X^{\otimes N}$ (NOT operator applied to all qubits) and  $c^{N}u_{1}$ (multi controlled phase rotation among $N$ qubits) to change the phase from the first state.

\begin{figure}[h]
\includegraphics[width=0.75\textwidth]{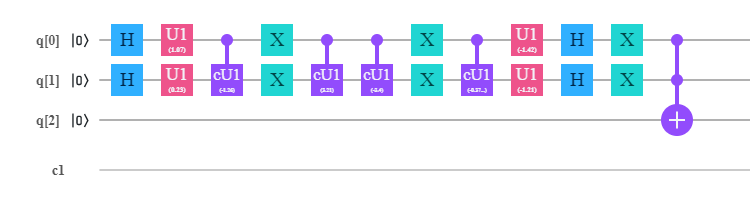}
\centering
\caption{Example of one circuit utilizing the HSGS algorithm.}
\label{TRES}
\end{figure}

Analogously to the rotation blocks, it is necessary to make the process in the `reverse' way for the weight operator. Aiming to take the weight state $\left| \psi_{w} \right\rangle $  to a balanced superposition state and then apply operators $H^{\otimes N}$ and $NOT^{\otimes N}$ in parallel. Figure \ref{TRES} illustrates a circuit built using the HSGS algorithm.

\subsection{Gradient descent algorithm}

To start the discussion about the gradient descent algorithm, it is necessary to have some functions clearly defined. Starting with the activation function, which is the mathematical function that maps our inputs and outputs. For any input and weight vectors:
\begin{equation}
\begin{split}
\left|\psi_{i}\right\rangle & =\left(\begin{array}{cccc}
e^{ix_{0}} & e^{ix_{1}} & \ldots & e^{i\alpha_{m-1}}\end{array}\right)^{T}\frac{1}{\sqrt{m}}\\
\left|\psi_{w}\right\rangle & =\left(\begin{array}{cccc}
e^{iw_{0}} & e^{iw_{1}} & \ldots & e^{iw_{m-1}}\end{array}\right)^{T}\frac{1}{\sqrt{m}}
\end{split}
\end{equation}
That is, the inner product is given by:
$$\left\langle \psi_{w}|\psi_{i}\right\rangle =\frac{1}{m}\sum_{j=1}^{m-1}e^{i\left(x_{j}-w_{j}\right)}=\frac{1}{m}\sum_{j=1}^{m-1}e^{i\alpha_{j}}$$
Where $\alpha_{j}=x_{j}-w_{j}$. So the output of the quantum circuit will be:
$$\left|\left\langle \psi_{w}|\psi_{i}\right\rangle \right|^{2}=\frac{1}{m^{2}}\sum_{j=1}^{m-1}\sum_{l=1}^{m-1}e^{i\left(\alpha_{j}-\alpha_{l}\right)}$$

Since the sine and cosine functions are odd and even, respectively, the symmetric terms will have the following property:
$$e^{i\left(\alpha_{j}-\alpha_{l}\right)}+e^{i\left(\alpha_{l}-\alpha_{j}\right)}=\cos\left(\alpha_{j}-\alpha_{l}\right)+\cos\left(\alpha_{l}-\alpha_{j}\right)=2\cos\left(\alpha_{j}-\alpha_{l}\right)$$
Thus, we may write in a generic way:
$$\left|\left\langle \psi_{w}|\psi_{i}\right\rangle \right|^{2}=\frac{1}{m^{2}}\sum_{j=1}^{m-1}\sum_{l=1}^{m-1}\cos\left(\alpha_{j}-\alpha_{l}\right)$$
With the activation function defined, the next step is to define the cost function. The cost function is a quantitative measure of the error that the neural network presents when classifying a training set. A common cost function for only one output may be defined as follows:
$$C=\frac{1}{2n}\sum_{k=1}^{n}\left(s_{k}-\left|\left\langle \psi_{w}|\psi_{i}\right\rangle \right|_{k}^{2}\right)^{2}$$
Where $n$ is the number of training inputs, $s_{k}$ is the desired output for a given input $k$ and $\left|\left\langle \psi_{w}|\psi_{i}\right\rangle \right|_{k}^{2}$ is the circuit output given the current weights. So $\left(s_{k}-\left|\left\langle \psi_{w}|\psi_{i}\right\rangle \right|_{k}^{2}\right)$ is the error for the input $k$. Defining  $\left(s_{k}-\left|\left\langle \psi_{w}|\psi_{i}\right\rangle \right|_{k}^{2}\right)=\left(\text{error}\right)_k$, it is possible to simplify the cost function as follows:
$$C=\frac{1}{2n}\sum_{k=1}^{n}\left(\text{error}\right)_{k}^{2}$$
Once the weights are updated, it is expected a lower cost function since it means that now the error is lower than with the old weights. The variation in cost function due to the variation of the weights is given by:
$$\Delta C=\frac{\partial C}{\partial w_{1}}\Delta w_{1}+\frac{\partial C}{\partial w_{2}}\Delta w_{2}$$
It may be rewritten through the gradient:
$$\Delta C=\overrightarrow{\nabla}_{w}C\cdot\Delta\overrightarrow{w}$$
On which $\overrightarrow{w}=\left(w_{1},w_{2}\right) $. Choosing a change in the weight vector as: $$\Delta\overrightarrow{w}=-\eta\overrightarrow{\nabla}_{w}C$$
We may guarantee that the cost function will be minimized, since:
$$\Delta C=-\eta\left\Vert \overrightarrow{\nabla}_{w} C\right\Vert ^{2}$$
Where $\eta$  is called learning rate. Summarily, the gradient descent algorithm basically consists of performing updates in the weight vector in the following way:

$$\overrightarrow{w}'=\overrightarrow{w}-\eta\overrightarrow{\nabla}_{w}C$$
Finally the cost function gradient has a more specific form:
$$\overrightarrow{\nabla}C=-\frac{1}{n}\sum_{k=1}^{n}\left(\text{error}\right)_{k}\overrightarrow{\nabla}_{w}\left(\left|\left\langle \psi_{w}|\psi_{i}\right\rangle \right|^{2}\right)$$
\section{Results}

All algorithms were implemented in Python using the NumPy library (\url{https://numpy.org/}) for analytical calculations and the Qiskit development kit (\url{https://qiskit.org/}) for Python, which provides not only simulators but also access to real devices in the IBM Quantum Experience cloud (\url{https://quantum-computing.ibm.com/}) . All tests on simulators and the quantum version of the perceptron were run from the cloud computing resource made available by Google, called Google Colab (\url{https://colab.research.google.com/}). Through this platform, Google provides an Intel(R) Xeon(R) CPU @ 2.30GHz with 2 cores and 12GB of RAM. Qiskit version 0.16.1 was also used in this environment, which has the following version of its modules:

\begin{itemize}
    \item qiskit-aer: 0.7.2,
    \item qiskit-aqua: 0.8.1,
    \item qiskit-ibmq-provider: 0.11.1,
    \item qiskit-ignis: 0.5.1,
    \item qiskit-terra: '0.16.1.
\end{itemize}

The environment provided by Google has its connection closed after 12 hours of duration. Thus, to execute the quantum version of the sigmoid it was necessary to use a personal computer to keep the execution for more time if needed. Therefore were run on a computer with an Intel(R) Core(TM) i5-6400 CPU @ 2.70GHz, 4 cores, and 8Gb of RAM. On the other hand, Qiskit version 0.19.6 was used on this computer with the following version of its modules:

\begin{itemize}
    \item qiskit-terra: 0.14.2, 
    \item qiskit-aer: 0.5.2, 
    \item qiskit-ignis: 0.3.3, 
    \item qiskit-ibmq-provider: 0.7.2, 
    \item qiskit-aqua: 0.7.3.
\end{itemize}

\subsection{Inner product}

To compare the results obtained with the proposed algorithms (rotation blocks and HSGS), the following quantitative measure of mean discrepancy, proposed by \citet{Tacchino}, was used:
\begin{equation}
D=\frac{\sum_{k=1}^{n}\left|O_{IBM}^{\left(k\right)}-O_{ideal}^{\left(k\right)}\right|}{n}
\end{equation}
where for $n$ calculated outputs, we make the difference between the ideal output ($O_{ideal}^{\left(k\right)}$  ), classically calculated, and the output of real IBM devices ($O_{IBM}^{\left(k\right)}$), obtained through Qiskit. Due to limitations imposed on the use of IBM hardware, such as utilization queues, we limit it to the case with $N=2$ qubits.  

Due to time limitations and the possibility of constructing infinite continuous vectors, it was chosen to work with only $8$ randomly generated vectors of continuous values. So that it becomes possible to combine the $8$ vectors to calculate $n=8^2=64$ different inner products. The code was run 5 times on 3 three different processors, 3 times it was only on \textit{ibmq\_vigo} as it is the best of the 3 processors compared to the analytic solution . The result is shown in Table \ref{tabela2}.

\begin{table}
\caption{Discrepancy for the inner product between continuous vectors}
\label{tabela2}
\centering
\begin{tabular}{llr}
\toprule
\cmidrule(r){1-3}
Processor & Rotation blocks & HSGS \\
\midrule
{ibmq\_5\_yorktown}    & 0.32855645449163706       & 0.338324540019847   \\ {ibmq\_valencia} & 0.1407079724761018 & 0.11474348544855864  \\ {ibmq\_vigo} & 0.10009663220763679 & 0.10272424045323603   \\ 
{ibmq\_vigo} & 0.11010601938883248 & 0.1821464251602708 \\
{ibmq\_vigo} & 0.1116498082272541 & 0.1082913676285323 \\
\bottomrule
\end{tabular}
\end{table}

The mean difference between the results of the algorithms was $\left(-1.10\pm3.27\right)\times10^{-2}$, that is, the standard deviation was greater than the mean in the module. This indicates that we can expect that depending on the execution, one algorithm or another may do a little better. This isn't what is expected when the HSGS was originally proposed by \citet{Tacchino}. A better investigation is needed, but one possibility of explanation is the improvement of the quantum processors structure - HSGS provides a significant advantage on processors that do not have native available multi-qubit operations - which reduces  the advantage of the HSGS algorithm.

\subsection{Evaluation of the binary case}

For comparison, we reproduced the binary calculus, where instead of using the operator $u_{1}$ responsible for performing a generic variation $\theta$ in the phase, an operator $Z$ is used that simply performs a change of signal.  We used 16 different unit and binary vectors made by $N=2$ qubits to calculate 256 different inner products, and executed 5 times in 3 different processors, the processor that exhibited the best results we executed 2 more times. The result can be seen in Table \ref{tabela1}.

\begin{table}
\caption{Discrepancy for the inner product between binary vectors}
\label{tabela1}
\centering
\begin{tabular}{llr}
\toprule
\cmidrule(r){1-3}
Processor & Rotation blocks & HSGS \\
\midrule
{ibmq\_5\_yorktown}    & 0.22016143798828125       & 0.22871780395507812   \\ 
{ibmq\_valencia} & 0.1013946533203125 & 0.1109771728515625  \\ 
{ibmq\_vigo} & 0.09790420532226562 & 0.08685302734375   \\ 
{ibmq\_vigo} & 0.09388351440429688 & 0.08945083618164062   \\ 
{ibmq\_vigo} & 0.10137939453125 & 0.0917205810546875   \\
\bottomrule
\end{tabular}
\end{table}

We found a result qualitatively equal to the previous case  with the mean difference between the discrepancy of the algorithms of $\left(1.40\pm8.83\right)\times10^{-3}$. To confirm that the proposed algorithms can reproduce results known in the literature, we used our algorithm for the inner product in the hybrid scheme proposed by  \citep{Tacchino} for the binary case. More details about the scheme can be read in the original work.

It was decided to work with a perceptron of 4 inputs that only require 2+1 qubit (considering the auxiliary qubit) and implement it in the \textit{ibmq\_vigo} device, which presented the better results in previous tests. The process proceeded as follows: First, a weight vector was randomly defined, and then from this vector, a training set was randomly constructed consisting of 5 vectors that result in a positive result and 50 that result in a negative output. After generating the training data, the weight vector was then `forgotten', and an initial weight vector was randomly generated to execute the training.

Thus, it is expected that during training the perceptron is able to recover the objective weight vector used in the construction of the training set. To quantify this learning process, a measure called affinity was used, which is the square of the inner product between the current weight vector and the objective, being then an affinity $1$ when the two vectors are identical. 

\begin{figure}[ht]
            \includegraphics[width=0.5\textwidth]{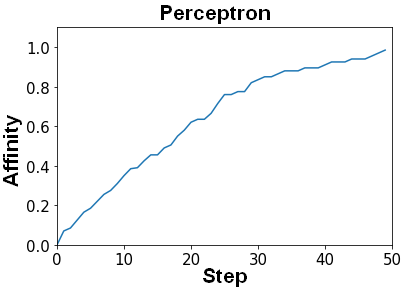}
            \centering
            \caption{Affinity $\times$ number of steps in the quantum device.}
            \label{pq}
 \end{figure}

It is also worth mentioning that at each new trial, the algorithm shuffles the order of the training data, training in a new random sequence. Finally, with the same training data set, the network was trained 59 times, restarting with an initial weight vector defined also randomly. The training also has a maximum limit of 50 steps for each execution. The average result for each step result is illustrated in Figure \ref{pq}. And we can see behavior that indicates that the perceptron is able to learn as expected.

\subsection{Simulation}
Considering that the sigmoid is a sensitive neuron model, and in order to test the feasibility of this model in quantum computers in the long term, to simulate the sigmoid it was chosen to use the simulator provided by Qiskit called `\textit{statevector \_simulator}'. This simulator does not include noise, it is a simulator of an ideal quantum circuit.

Looking at the activation function we can see that the inputs (and weights) always appear in pairs as arguments of trigonometric functions, so we actually have $m-1$ free parameters for $N$ qubits and $m=2^N$ states. Which makes it possible to represent the same result with just $m-1$ entries (and weights). From these considerations, a neuron with 3 inputs (2+1 qubits) was simulated.

Following a procedure as discussed in the last section, first an ideal objective weight vector of 3 randomly defined components was generated, where the components vary between $0$ and $2\pi$. These values are then encoded as the phases of the state vector amplitude coefficients of the system.

From this, a training set with 200 randomly defined input vectors is generated. The ideal output for each input is then calculated and, after this step, the objective weight vector is `forgotten' and a new initial weight vector is generated randomly.

\begin{figure}[ht]
\includegraphics[width=0.5\textwidth]{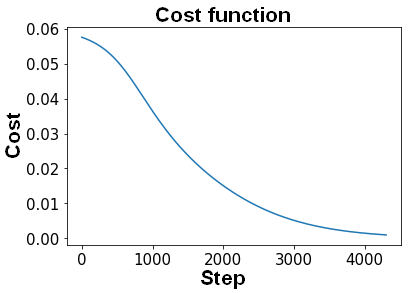}
\centering
\caption{Behavior of the cost function for the sigmoid in the simulator. }
\label{img2}
\end{figure}

A learning rate of $\nu=0.1$ was adopted and two stopping conditions were defined: The training should end if the cost function is reduced to less than $10^{-3}$ or if it start to increase. After approximately 4000 steps, the cost function has reduced from an initial value of approximately $5.75\times10^{-2}$ to approximately $1.74 \%$ of its initial value, i.e. $1.00\times10^{-3}$. This behavior can be seen in Figure \ref{img2}.

As we can see, the cost function decreases as expected, which implies that the sigmoid neuron is learning from the training set. We can also observe that with the set of parameters that we run the neural network, the cost function appears to stabilize at $1.00\times10^{-3}$, not decreasing to values smaller than this. This agrees with the fact that the run ended when the cost function was increased. One hypothesis is that it would begin to oscillate around this value. If we want to further decrease the cost function, we can think that we could change the learning rate, change the training dataset, or even use a different learning algorithm.

\subsection{Implementation}

Considering that it is a naturally sensitive model to noise, it was thought to use a quantum device that would perform better, despite the long queues. To compensate, the training set was reduced to 50 inputs and the device \textit{ibmq\_santiago} was used. Quantum volume, short for QV (\textit{quantum volume}), is a metric that IBM has defined to measure the performance of its real quantum computers. Therefore, we can see it as quantification of this performance, and the higher the QV, the better the performance of the device. Two of the three devices used so far (\textit{ibmq\_valencia } and \textit{ibmq\_vigo }) have $QV=16$, while \textit{ibmq\_5\_yorktown} has $ QV=8$. The choice of \textit{ibmq\_santiago} was due to having $QV=32$.

\begin{figure}[ht]
\includegraphics[width=0.5\textwidth]{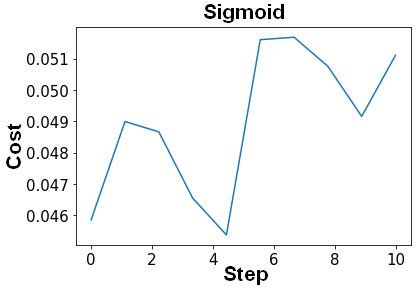}
\centering
\caption{Behavior of the cost function for the sigmoid on the device `\textit{ibmq\_santiago}'. }
\label{sq}
\end{figure}

\begin{figure}[ht]
\includegraphics[width=0.5\textwidth]{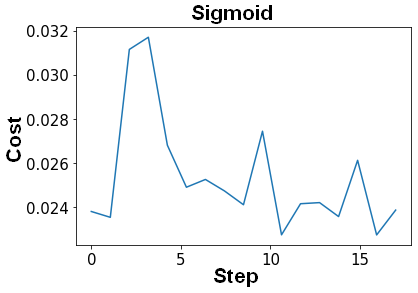}
\centering
\caption{Behavior of the cost function for the sigmoid on device `\textit{ibmq\_5\_yorktown}'. }
\label{sq2}
\end{figure}

That said, it was necessary to remove the stopping condition for training based simply on the increase in the cost function since it presented an oscillation from the beginning. The result can be seen in Figure \ref{sq}. At the same time, the same code was executed in another device, \textit{ibmq\_5\_yorktown} ($QV=8$), due to smaller queues. The obtained results can be checked out in Figure \ref{sq2}.


Graphically, we can see that there is an oscillatory behavior. And calculating the average of the steps that were possible to be performed in real quantum devices, $10$ in \textit{ibmq\_santiago} and 17 in \textit{ibmq\_5\_yorktown}, we have that the cost value was in average $\left(4,90\pm0,23\right)\times10^{-2}$ and $\left(2,54\pm0,25\right)\times10^{-2}$, approximately the same initial values ($4.59\times10^{-2}$ and $2.38\times10^{-2}$). That is, although we show that the model works in an ideal quantum device, it did not appear to be capable of learning in a real quantum device. The most likely hypothesis is that it is due to noise. In order to try to improve this model targeting current or short-term devices, we could think about some hypotheses similar to the one discussed how to improve the model implemented in the simulator, that is, change the learning rate, the training dataset, or even the learning algorithm implemented.

Another point is that while the model in the simulator had approximately 4,000 steps, in the quantum device we had time to perform only less than $20$ due to the high number of calculations that need to be performed on a quantum device and IBM's access policies to its public devices.  However, it is worth remembering that at no time did the cost function in the simulator increase in these 4,000 steps, while in the real quantum device it seems to be fluctuating around the initial value since the beginning, which is an expected characteristic in view of the noise problems that the current devices still face since the sigmoid model was built to be sensitive to small changes. However, the possibility remains open to allow the sigmoid to run longer to verify if at some point there is a tendency for the neuron to be learning, even with the noise present in current devices.

\section{Conclusion}

Therefore, synthesizing the results obtained during this work, based on a 2019 article \citet{Tacchino}, two algorithms were implemented to calculate the inner product between vectors in a quantum computer. However, differing from the results obtained in the literature, it was not possible to verify a significant difference between the results obtained by each algorithm, that is, the discrepancies obtained when calculating the same inner product in the same device, but using the two different algorithms, did not show a significant difference between each other. Among the hypotheses raised to explain this discrepancy, the main one is the architecture of quantum processors.

For the proposed hybrid machine learning model, when implemented in the simulator provided by Qiskit, it presented satisfactory results that demonstrate a good potential in terms of learning capacity. As for the implementation in real devices, there are still a lot of challenges to make use of real quantum devices: a significant amount of noise, limitations in the number of qubits, the time needed to submit the circuit and obtain the results, among others.

This result sheds light on some important questions. The first of a technical nature, comparing the results obtained between simulators and real devices, shows us the importance of the noise mitigation, both the development of theoretical tools and the development of better hardware since it is coherent to assume that the only reason for the model to have such discrepant results between the simulator and the real device, it is due to the presence of a significant amount of noise in current real quantum devices.

And the second question highlights the limitations that are imposed by using a free public version of a product offered by a large international company. The possibility of having easy access to the most modern current technological devices, in this case, real quantum devices, is shown to be a possible great differential for the research. Not only for this present research, but it's natural to think how significant this is even for national scientific and technological development. We could still see how this is a fast-growing area, as we could see how the advantage that one algorithm had over another was largely suppressed due to technological development in just a few couples of years.

In conclusion, we believe that the work represents another step on the right path in the development of quantum artificial intelligence.



\bibliographystyle{cen}
\bibliography{exemplo_bib_cen}

\end{document}